\documentclass{IEEEtran}

\usepackage{setspace}
\usepackage{mathptmx}
\usepackage{amsmath}
\usepackage{amssymb}
\usepackage{epsfig}
\usepackage{amsthm}
\usepackage{amsfonts}
\usepackage{subfigure}
\usepackage{stmaryrd}
\usepackage[dvips]{color}
\usepackage{graphicx}
\usepackage{mathcomSTEP}

\allowdisplaybreaks

\theoremstyle{remark}
\newtheorem{rem}[thm]{Remark}

\newtheorem{definition}{Definition}

\begin{document}

\title{
On Capacity of the Dirty Paper Channel with Fading Dirt in the Strong Fading Regime
}

\author{%
\authorblockN{%
Stefano~Rini\authorrefmark{1} and Shlomo~Shamai~(Shitz)\authorrefmark{2} \\
}

\authorblockA{%
\authorrefmark{1}
National Chiao-Tung University, Hsinchu, Taiwan\\
E-mail: \texttt{stefano@nctu.edu.tw} }

\authorblockA{%
\authorrefmark{2}
Technion-Israel Institute of Technology,  Haifa, Israel \\
E-mail: \texttt{sshlomo@ee.technion.ac.il} }

\thanks{
%
The work of S. Shamai was supported by the Israel Science Foundation (ISF) and by the European FP7 NEWCOM\#.
%
}
}

\maketitle

\begin{abstract}
The classical ``writing on dirty paper'' capacity result establishes that full interference pre-cancellation can be attained in Gel'fand-Pinsker problem
with additive state and additive white Gaussian noise.
This result holds under the idealized assumption that perfect channel knowledge is available at both transmitter and receiver.
While channel knowledge at the receiver can be obtained through pilot tones, transmitter channel knowledge is harder to acquire. 
For this reason, we are interested in characterizing the capacity under the more realistic assumption that only partial channel knowledge is available at the transmitter.
We study, more specifically, the dirty paper channel in which the interference sequence in multiplied by fading value unknown to the transmitter but known at the receiver.
For this model, we establish an approximate characterization of capacity for the case in which fading values vary greatly in between channel realizations.
In this regime, which we term the ``strong fading'' regime,  
the capacity pre-log factor is equal to the inverse of the number of possible
fading realizations.
\end{abstract}

\section{Introduction}
The capacity of the classical the Gel'fand-Pinsker (GP) problem \cite{GelFandPinskerClassic} characterizes the limiting performance of interference pre-cancellation.
In this model, the channel outputs are obtained as a random function of the channel inputs and a state sequence; the state sequence is provided anti-causally
to the encoder but is unknown at the decoder.
This channel model is motivated by many practical downlink networks in which the transmitter wishes to simultaneously communicate to multiple receivers:
in this scenario the codeword of one user can be treated as known interference when coding for of another user.
The ``writing on dirty paper'' capacity result from Costa \cite{costa1983writing} establishes that the presence of the state sequence does not reduce
capacity in the Gaussian version of the GP problem, regardless of the distribution or power of this sequence.
%
Although very promising, this result is not easily translated into practical transmission strategies since it assumes perfect channel knowledge at both the 
transmitter and the receiver. 
Receiver channel knowledge can be acquired through various channel estimation strategies,  pilot tones and channel sounding.
This channel knowledge is successively made available at the transmitter through feedback.
In many communication environments, the channel conditions vary through time and obtaining a reliable channel estimate at the transmitter is very costly.
For this scenario, we wish to characterize the limiting performance of interference pre-cancellation when only partial transmitter channel knowledge is available.
%

The first channel model which address the impact of transmitter channel knowledge in the GP problem is the ``writing on fading dirt''
channel \cite{zhang2007writing}.
This model is a variation of the ``writing on dirty paper'' channel of \cite{costa1983writing} in which interference sequence
is multiplied by a fading value which is made available at the receiver.
%
%
The capacity of this model is a special case of the channel in  \cite{cover2002duality}, an extension of the GP problem in which partial state information is
provided to both transmitter and receiver.
The capacity of the channel in \cite{cover2002duality} is expressed as a maximization over the distribution of an auxiliary random variable which cannot be easily determined.
For this reason, neither closed form expressions nor numerical evaluations of capacity are known.
Outer and inner bounds to the capacity are derived in \cite{grover2007writing} while
achievable rates under Gaussian signaling and lattice strategies are derived in \cite{avner2010dirty}.
%
%
The  fading dirty paper channel in which the fading values is constant trough successive channel uses is studied in \cite{piantanida2009capacity}.
Fundamental bounding techniques for this channel are drawn from the ``carbon copying onto dirty paper'' \cite{LapidothCarbonCopying}.

In the following, we focus on the writing on fading dirt problem for the slow fading case and consider the case in which the state sequence
is an iid Gaussian sequence.
For this channel we determine inner and outer bound to capacity which depend on the number of possible fading realizations.
We consider, more specifically, the case in which the fading takes two different values and the case in which it takes $M$ possible values.
For the case in which fading takes two possible values, we obtain a characterization of capacity to within a gap which depends the distance
between fading realizations.
For the case  of $M$ possible realizations, instead, we characterize capacity in a regime where the fading realization are exponentially spaced apart.
In this regime, which we term the ``strong fading regime'',  the pre-log of the capacity is equal the inverse of the number of fading realizations.

The remainder of the paper is organized as follows: in Sec. \ref{sec:DPC-SF-RCSI} we introduce the channel model.
Sec. \ref{sec:Related Results} presents related results.
In Sec. \ref{sec:Two value} we study the case of two possible fading realizations while,
in Sec. \ref{sec:M values} the case of $M$ realizations.
Finally, Sec. \ref{sec:Conclusion} concludes the paper.

\noindent
\underline{If omitted, proofs can be found in the appendix or in the }
\underline{bibliographic reference.}

%

\section{Channel Model}
\label{sec:DPC-SF-RCSI}

\begin{figure}
\centering
\includegraphics[width=.5 \textwidth]{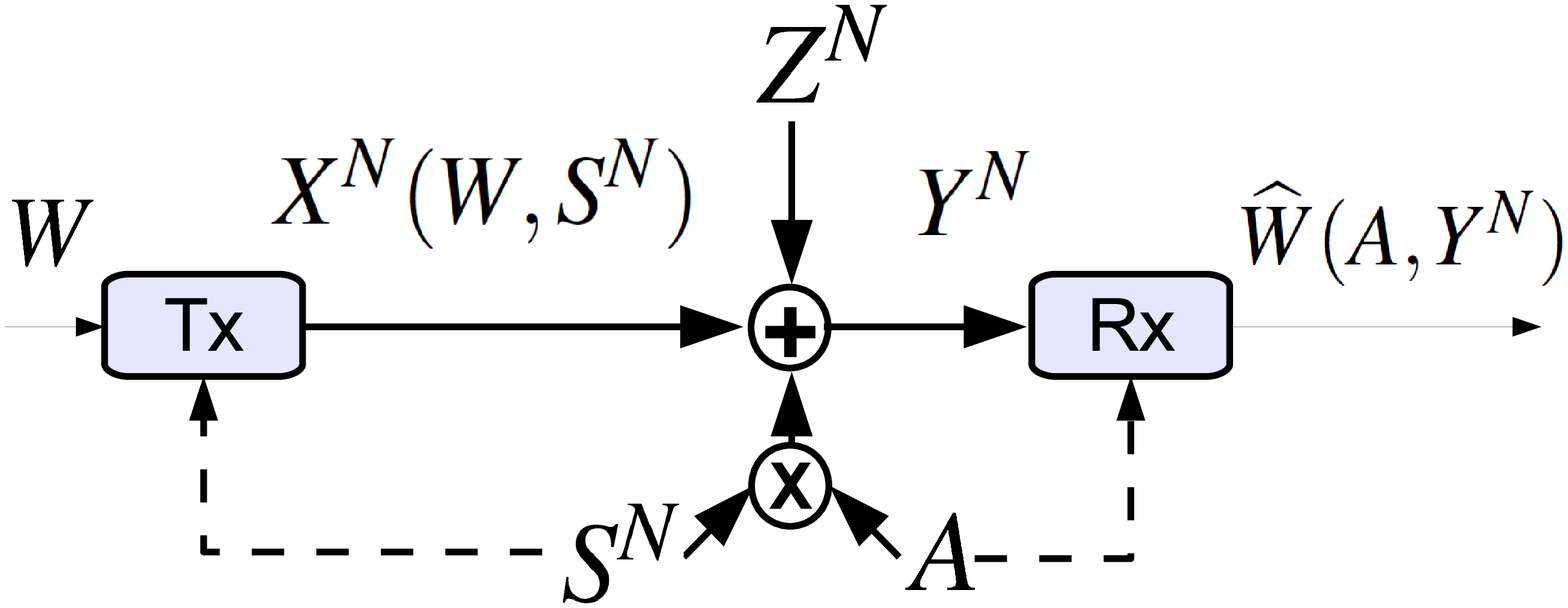}
\vspace{-4 cm}
\caption{The Dirty Paper Channel with Slow Fading and Receiver Channel State Information (DPC-SF-RCSI).}
\vspace{-.3 cm}
\label{fig:DPC-SF-RCSI}
\end{figure}

The Dirty Paper Channel with Slow Fading and Receiver Channel State Information (DPC-SF-RCSI) in Fig. \ref{fig:DPC-SF-RCSI},
is defined by the input/output relationship
\ea{
Y^N = X^N + A S^N + Z^N,
\label{eq:DPC-SF-RCSI}
}
where $Z_i \sim \Ncal (0,1),  \ i \in [1 \ldots N]$ and iid.
The channel input $X^N$ is subject to an average second moment constraint:
\ea{
\Ebb \lsb \sum_{i=1}^N X_i^2 \rsb \leq N P.
}
The state sequence $S^N$ is iid, Gaussian distributed  with zero mean and unitary variance and is provided to the encoder but not to the decoder.
The fading values $A$ is chosen from a set $\Acal=\{a_1 \ldots a_M\}, \ 0\leq  a_1 < a_2 \ldots < a_M $, fixed before transmission and made available 
at the decoder but not at the encoder.

A $(2^{NR},N,P_e)$ code for the DPC-SF-RCSI is defined by an encoding function $X^N(W,S^N)$, a decoding function $\Wh(Y^N,A)$ and an error probability 
\ean{
P_e = \f 1 {2^{NR} M } \sum_{w=1}^{2^{NR}} \sum_{j=1}^M P \lsb \Wh(Y^N,a_j) \neq  W| X^N(w,S^N) {\rm \ was \ sent} \rsb.
}
A rate $R$ is achievable if, for any $\ep>0$, there exists a code $(2^{NR'},N,P_e)$ such that $R'\geq R$ while $P_e \leq \ep$.
%
%

\begin{figure}
\centering
\includegraphics[width=.45 \textwidth]{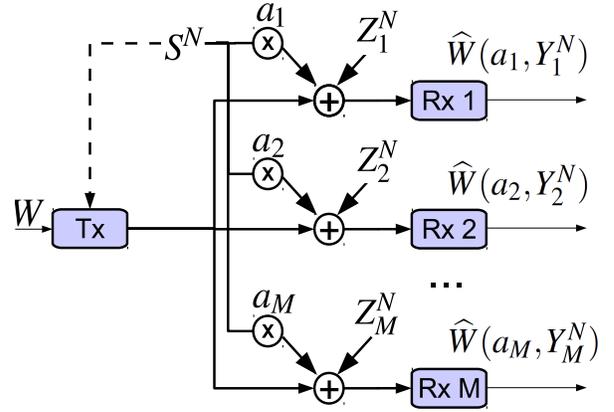}
\vspace{-1 cm}
\caption{The compound broadcast channel equivalent to the DPC-SF-RCSI with $M$ possible fading values.}
\label{fig:compoundChannel}
\vspace{-.5 cm}
\end{figure}

We consider the compound channel approach to the capacity of the DPC-SF-RCSI, as originally proposed by Shannon \cite{shannon1958channels}.
In this approach, the capacity of the channel model in \eqref{eq:DPC-SF-RCSI} is equivalent to the capacity of the compound broadcast channel
in Fig. \ref{fig:compoundChannel} for which
\ea{
Y_{j}^N=X^N+ a_j S^N + Z_{j}^N \quad a_j \in  \Acal
\label{eq:compoundChannel}
}
%
%
%

Finally, we present a more general class of models which is equivalent to that in \eqref{eq:DPC-SF-RCSI}.

\begin{rem}{\bf Generalized channel model.}
\label{rem:Generalized Channel Model}
The channel model in \eqref{eq:DPC-SF-RCSI} is equivalent to any channel model where the input output relationship is described as
\ea{
Y_i = X_i + A' S'_i + Z'_i,
\label{eq:generalized channel}
}
for $S' \sim \Ncal(0,Q')$, $Z' \sim \Ncal(0,N')$ and $\sum_{i=1}^N\Ebb[X_i^2] \leq N P'$.
\end{rem}

\section{Related Results}
\label{sec:Related Results}
We shall next review some of the results in the literature related to the DPC-SF-RCSI.

\smallskip
\noindent
\underline{\bf Dirty paper channel with fading dirt:}
when the fading values in \eqref{eq:DPC-SF-RCSI} change at each channel use, that is when the channel output is obtained as
\ea{
Y^N = X^N + A^N S^N + Z^N,
\label{eq:DPC-FD}
}
for some iid sequence  $A^N$ drawn from the distribution $p_A$, the channel is termed ``dirty paper channel with fading dirt'' \cite{avner2010dirty}.
\begin{thm}{\bf Capacity of the dirty paper channel with fading dirt \cite[Th. 1]{cover2002duality}.}
\label{th:Capacity of the ergodic DPC-PF}
The capacity of the channel in \eqref{eq:DPC-FD} is obtained as
\ea{
C=\max_{P_{U,X|S}} \ I(Y; U| A) - I(U;S),
\label{eq: capacityWritingOnFadingDirt QS}
}
\end{thm}
Equation \eqref{eq: capacityWritingOnFadingDirt QS} contains a maximization over $P_{U,X|S}$ which cannot be easily evaluated explicitly.

\smallskip
\noindent
\underline{\bf Carbon copying onto dirty paper:} when the state sequence is different in each fading realization, we obtain the carbon copying paper model in \cite{LapidothCarbonCopying}, that is
\ea{
Y_j^N = X^N + a S_j^N + Z_j^N,
\label{eq:DPC-SF-RCSI}
}
for $j \in [1 \ldots M]$ and where $S_j^N$ is an iid Gaussian sequences with zero mean and unitary variance for each $j$.
In \cite{LapidothCarbonCopying} inner and outer bound to the capacity region are derived but capacity has yet not been determined.
The next outer bound is be relevant in deriving an outer bound for the  DPC-SF-RCSI.
\begin{thm} {\bf Outer bound for the 2 user case  \cite[Th. 3]{LapidothCarbonCopying}.}
%
The capacity $C$ of the 2 user carbon copying channel is upper bounded as
\ea{
C \leq R^{\rm OUT} =
\lcb
\p{
\f 1 2 \log \lb \f{1+P+a^2+2a \sqrt{P}}{1+a^2/2} \rb  &  a^2 \leq 2 \\
\f 1 2 \log \lb \f{1+P+a^2+2 a \sqrt{P}}{a/\sqrt{2}} \rb \\
\ \ \quad - \lsb \f 14  \log \lb \f {a^2}{2P+2} \rb \rsb^+   &  a^2 >  2. \\
}
\rnone
}
\end{thm}
A simple inner bound for the two user case can be derived using binning and Gaussian signaling as in the original DPC channel.
In this scheme, the channel input is comprised of two codewords: one codeword is pre-coded against the state sequences
while another codeword treats the states as noise.

\begin{thm}{\bf Inner bound for the  2 user case \cite[Th. 4]{LapidothCarbonCopying}. }
\label{th:Interference as Noise + Binning Inner Bound}
The capacity $C$ of the 2 user carbon copying channel is lower bounded as
\ea{
& C \geq R^{\rm IN} =\lcb
\p  {
\f 1 2 \log \lb 1 + \f P {a^2/2+1} \rb  & a^2/2 < 1 \\
\f 1 2 \log \lb \f{P+a^2/2+1 }{a^2} \rb \\
 \quad + \f 14 \log \lb \f {a^2}2 \rb  & 1 \leq a^2/2 \leq P+1 \\
\f 1 4 \log(1 + P)                      &  a^2/2 \geq  P + 1.
}
\rnone
\label{eq:Interference as Noise + Binning Inner Bound}
}
\end{thm}

%
\smallskip
\noindent
\underline{\bf DPC-SF-RCSI:} the capacity of the DPC-SF-RCSI is unknown in general but it must reduce to the capacity of the classical DPC channel when the fading
realizations all approach the same value.
%
In this limit, Costa pre-coding must be optimal: the next lemma describes how the performance of Costa pre-coding decreases as the distance
between fading realizations increases.

\begin{lem}{\bf Constant additive gap for $M$, small fading values.}
\label{lem:Constant additive gap for M, small fading values}
An outer bound to the capacity of the $M$ user DPC-SF-RCSI is
\ea{
R^{\rm OUT} = \f 12 \log(1+P),
\label{eq:trivial outer bound M}
}
and it can be attained to within the gap $G$ by Costa pre-coding, for
\ea{
G=R^{\rm OUT}-R^{\rm IN}\leq  \f 12 \log \lb 1+  \f {\ep^2 P}{P+a_1^2+1}  \rb,
}
and with $\ep = a_M-a_1$.
\end{lem}
Lem. \ref{lem:Constant additive gap for M, small fading values} provides a tight characterization of capacity when $a_M\approx c(a_1 + 1/P)$ for some small constant $c$.

\section{Channel with two fading realizations}
\label{sec:Two value}

In this section we derive an outer bound inspired by a bounding technique in \cite[Th. 3]{LapidothCarbonCopying} and an inner
bound similar to that of \cite[Th. 4]{LapidothCarbonCopying} and then determined the gap between these two bounds.
In the derivation of the outer bound, an important observation is that the capacity of the channel is decreasing in the variance of the sequence $A S^N$.

\begin{rem}{\bf Capacity decreases as fading increases.}
\label{rem:capacity decreasing scaling fading}
The capacity of the channel
\ea{
Y^N=X^N+ \gamma A S^N + Z^N
}
for $\gamma \in [0,1]$ is decreasing in $\gamma$.
\end{rem}
With this result, we can establish a tighter outer bound than the one in \cite[Th. 3]{LapidothCarbonCopying}.
\begin{thm}{\bf Outer bound for two fading values.}
\label{th:Outer bound 2}
%
The capacity $C$ of the 2 user DPC-SF-RCSI  is upper bounded as
\ean{
& C \leq R^{\rm OUT}  = \min_{ \gamma \in [0,1]} \lcb \f 1 2 \log\lb 1+P+\gamma^2 a_1^2 + 2 \gamma a_1 \sqrt{P}  \rb \rnone \\
& \quad  \lnone +\f 12 \log \lb 1+P+\gamma^2 a_2^2 + 2 \gamma a_2 \sqrt{P} \rb -\f 1 2 \log(\gamma^2 (a_2-a_1)^2 ) \rcb
}
for which
\ea{
  R^{\rm OUT} & \leq \lcb \p{
  \f 14 \log (1+P) + \f 14 \log\lb P+1 +a_2^2 \rb  & a_2 a_1 \geq P+1  \\
  \quad -\f 14 \log (a_2-a_1)^2+ \f 3 2  & a_2 a_1 <P+1 
  } \rnone
\label{eq:Outer bound 2 fading}
}
\end{thm}

\begin{IEEEproof}
The first step of the outer bound is similar to the bounding in \cite[Th. 3]{LapidothCarbonCopying} while the second passage makes use of the observation in
Rem. \ref{rem:capacity decreasing scaling fading}.
\end{IEEEproof}
%
%
%
%
%
Drawing from the results available for the classical GP problem, an effective transmission strategy is the following.
The channel input is divided in two codewords: one codeword is pre-coded against a realization of the state while the other codeword treats the state as noise.
For the codeword which pre-codes against the state, the transmitter considers two transmission slots:
in the first slot it pre-codes against the sequence $a_1 S$ while in the second slot it pre-codes against the sequence $a_2 S$.
%
%
The next theorem determines the achievable rate of this strategy.
\begin{thm} {\bf Inner bound for two fading values.}
\label{th:inner bound for two fading values}
The capacity $C$ of the 2 user DPC-SF-RCSI is lower bounded as
\ea{
\Ccal \geq R^{IN} =
&\lcb \p{
\f 12 \log \lb 1  + \f P {1 + a_2^2} \rb                    & a_2^2 < 1  \\
\f 12 \log \lb 1+ P+ a_2^2 \rb \\
\quad - \f 14 \log (a_2^2)-1/2  & 1< a_2^2 < P+1 \\
\f 1 4 \log \lb 1+ P\rb                                   & a_2^2 \geq  P+1   \\
}
\rnone
\label{eq:inner bound for two fading values}
}
%
\end{thm}
\begin{IEEEproof}
Consider the transmission scheme in which the channel input, $X$, is comprised of two codewords:
(i) $X_{N}^N$ ($N$ as in ``\emph{state as Noise}'') which treats the state as noise and
(ii) $X_{P}^N$ ($P$ as in ``\emph{Pre-coded against the state}'')  is pre-coded against the sequence $S'=[S^{N/2} \  a_2 S_{N/2+1}^N]$.
The codewords are Gaussian  distributed: $X_{N}^N$ has power $\al P$ while $X_{P}$ has power $\alb P$ for $\al \in [0,1], \ \alb=1-\al$.
Once the codeword $X_{N}^N$ has been decoded, it is removed from the channel output and, successively, the codeword $X_{P}^N$ is decoded.
For the first $N/2$ transmission, the codeword pre-codes against the interference experience at the first receiver, while for the second $N/2$ transmissions
for the interference experienced at the second receiver.
This strategy achieves the rate
\ea{
R^{\rm IN}(\al) =
&   \f 12 \log \lb 1 + \f{\al P} {1 +\alb P+ a_2^2Q} \rb + \f 14 \log \lb 1 + \alb P  \rb
\label{eq:Interference as Noise + Binning Inner Bound} \\
& \ \ +\f 14  \log \lb \max \lcb 1, \f {(\alb P+1)(\alb P+a_2^2 Q+1) }
{\alb P+2a_2^2 Q \alb P +a_2^2 Q+1}\rcb \rb,
\nonumber
}
for any $\al \in [0,1]$.
The optimization over $\al$ yields the rate in \eqref{eq:inner bound for two fading values}.
\end{IEEEproof}
We next establish a gap between inner and outer bound as a function of the two fading realizations.
\begin{thm}{\bf Approximate capacity for two fading values.}
\label{th:Approximate Capacity 2}
%
The capacity $C$ of the 2 user carbon copying channel can be attained to within a gap of $G$ defined as
\ea{
G=R^{\rm OUT}-R^{\rm IN} \leq \f 12 \log \lb  \f{a_2+a_1}{a_2-a_1}\rb + 2   
\label{eq:gap 2 fading}
}
\end{thm}
\begin{IEEEproof}
The expression in \eqref{eq:gap 2 fading} is obtained by comparing the expression in \eqref{eq:inner bound for two fading values} and the expression
in \eqref{eq:Outer bound 2 fading} for the cases (i) $a_1 a_2 \geq P+1$, $a_2> \sqrt{P+1}$ (ii) $1< a_2^2 < P+1$ and (iii) $a_1 a_2 < P+1$, $a_2> \sqrt{P+1}$.
\end{IEEEproof}

The result in Th. \ref{th:Approximate Capacity 2} well characterizes the capacity when there the two fading realizations are sufficiently separated.
Consider, in particular, the scenario in which $a_2^2=c a_1^2$ for $c>1$, then the gap scales as
\ea{
\f 12 \log \lb \f {c+1}{c-1} \rb =\f 12 \log\lb 1+\f 2 {c-1}\rb
}
which is less than $1$ when $c \geq 2$.
\bigskip

Together with the result in Lem. \ref{lem:Constant additive gap for M, small fading values}, we conclude that the capacity for the case where $a_2=a_1 c$ is
well characterized for all values of $c$, for instance
\ea{
c\geq 2, \ c\leq  1 + \f 3 P + \f 3 {a_1}  \implies G \leq 1
}
%

We also notice that when $a_2^2 \geq  P+1$ the capacity of the channel is approximatively half of the capacity of the channel without fading.
In the next section we generalize this result to the case where the fading takes $M$ possible values.
That is, we determine a regime in which the capacity of the channel is $M^{-1}$ times the capacity of the channel without fading.

\section{Channel with $M$ fading realizations}
\label{sec:M values}

The fundamental outer bounding technique in characterizing the capacity of the two fading channel was first introduced in \cite{LapidothCarbonCopying}.
%
%
We now present a further refitment in the outer bound which makes it possible to determine the approximate capacity of the channel in a subset of the
 parameter regime.
In this regime the fading realization are greatly spaced apart and the capacity of the channel is approximatively a fraction $M$ of the capacity of the channel
without fading.
%
%
%
We term this regime as ``Strong fading'' regime.

\begin{definition}{\bf ``Strong fading'' regime.}
\label{def: strong fading}
A  DPC-SF-RCSI with $M$ fading realizations is said to be in the ``strong fading'' regime if
\ea{
a_1^2  = 0,  \quad a_j^2 \geq (P+1)\sum_{q=1}^{j-1} a_{q}
\label{eq: strong fading}
}
\end{definition}
From the definition in Def. \ref{def: strong fading} we have that the fading coefficients for $j\geq 2$ grow approximatively exponentially spaced, as in a geometric series.
\begin{thm}{\bf Outer bound for $M$ fading values in the ``strong fading'' regime.}
\label{th:Outer bound M}
The capacity $C$ of the M user DPC-SF-RCSI in the ``strong fading'' regime is upper bounded as
\ea{
C \leq R^{\rm OUT} & = \f 1 {2M} \log(1+P) + 3+\f {\log M} M
\label{eq:outer bound M}
}
\end{thm}

\begin{IEEEproof}
The key in deriving this novel outer bound is the careful choice of side information provided at each receiver in the compound channel.
Using Fano's inequality we have
\ean{
N(R -\ep)& \leq  \min_{j \in \{1 \ldots M\}} \lcb I(Y_j^N; W) \rcb \leq \f 1 M \lb \sum_{j=1}^M I(Y_j^N; W) \rb
}
For each term $I(Y_j^N; W)$ we provide the side information $V_j^N$ defined as
\ean{
V_1^N & =V_2^N=\emptyset \\
V_j^N & = \lsb V_{j-1}^N \ S_{j-1}^N + Z_{j-1}^N-Z_1^N \rsb \quad j \geq 3.
}
With this side information we write
\ean{
\sum_{j=1}^M I(Y_j^N; W) & \leq
\sum_{j=1}^M I(Y_j^N ; W| V_j^N)  \\
& = \sum_{j=1}^M \lb  H(Y_j^N | V_j^N)-H(Y_j^N | W, V_j^N) \rb
\label{eq:pos minus neg }
}
For the positive entropy terms have
\ean{
H(Y_q^N | V_q^N)
& = H(X+a_q S + Z_q | a_2 S +Z_2, a_3 S + Z_3 \ldots a_{j-1} S + Z_q) \\
& \leq H(X+a_q S + Z_q | S + \Zt)\\
& \leq \f12 \log 2 \pi e \lb (\sqrt{P}+a_q)^2 - \f {(\sqrt{P}+a_q)^2 a_{j-1}^2}{a_{j-1}^2+1}\rb
}
where, in the last passage, we have used the maximal ratio combining principle with
\ea{
\Zt = \f {\sum_{j=1}^{q-1} a_j Z_j}{\sum_{j=1}^{q-1} a_j^2} \sim \Ncal \lb 0, \f {1} {\sum_{j=1}^{q-1} a_j^2} \rb
}
For the negative entropy terms we let
\ea{
T_q = -\sum_{j=1}^{q-1} H(Y_j^N| W, V_j^N)
}
and establish a recursion of the form
\ean{
T_q & = T_{q-1} - H(Y_j | W , a_2 S^N + Z_2^N - Z_1^N, \ldots a_{j-1} S^N+Z_{j-1}^N) \\
& \leq  - \sum_{j=1}^{q-1} H(V_j^N |W, V_{j-1}^N)-H(X^N-Z_1^N|V_{q}).
}
Each term $H(V_j^N | W, V_{j-1}^N)$ for $j\geq 3$ can be bounded as
\ean{
& H(V_j^N | W, V_{j-1}^N) \\
& = H(X+a_q S + Z_q | a_2 S +Z_2, a_3 S + Z_3, \ldots, a_{j-1} S + Z_{j-1}) \\
& \leq H(X+a_q S + Z_q | S + \Zt)\\
& \leq \f12 \log 2 \pi e \lb (\sqrt{P}+a_q)^2 - \f {(\sqrt{P}+a_q)^2 a_{j-1}^2}{a_{j-1}^2+1}\rb \\
}
where we have again used the maximal ratio combining principle.
%
The difference $H(Y_j^N|V_j)-H(V_j^N | V_{j-1}^N)$ for $j\geq 2$ is bounded as
\ean{
&  H(Y_j^N|V_j)-H(V_j^N | V_{j-1}^N)  \leq \f 12 \log \lb \f{ P+1 + \f {a_j^2}{\sum_{j=1}^{q-1} {a_j^2}+1}} {1+ \f{a_j^2}{1+ \sum_{j=1}^N a_j^2} }\rb  \\
& \quad \quad  = \f 12 \log \lb \f{ (\sum_{j=1}^{q-1} a_j^2+1)(P+1) + a_j^2} {\sum_{j=1}^{q} a_j^2+1} \rb.
}
When  condition \eqref{eq: strong fading} holds, we have therefore have
\ean{
 H(Y_j^N|V_j)-H(V_j^N | V_{j-1}^N) \leq \f 12 \log \lb 1+\f{3a_q^2} {\sum_{j=1}^{q} a_j^2+1} \rb \leq 2.
}
The only remaining term is $H(Y_1)-H(Y_1|V_M)$ for which we have
\ean{
& H(Y_1)-H(Y_1|V_M) \leq H(Y_1)-H(Y_1|V_M, S^N,X^N) \\
& = \f 12 \log(1+P) +\log M.
}
\end{IEEEproof}
A simple inner bound can be obtained by having the encoder pre-codes  against the term $a_j S_i, \ j \in[1 \ldots M]$ for a portion $1/M$ of the time.

\begin{thm}{\bf Inner bound for $M$ fading values.}
\label{th:Inner bound M}
The capacity $C$ of the $M$ user DPC-SF-RCSI is lower bounded as
\ea{
C \leq R^{\rm IN} & = \f 1 {2M} \log(1+P)
\label{eq:Inner bound M}
}
\end{thm}
\begin{IEEEproof}
As in the proof of Th. \ref{th:inner bound for two fading values}, consider the codeword $X_P^N$  pre-coded against the interference sequence $\St^N$ where
\ea{
\St_{jN/M}^{(j+1)N/M} = a_j S_{j N/M}^{(j+1)N/M}.
}
This achieves full interference pre-cancellation at receiver $j$ for the transmission $[j N/M \ldots (j+1) N/M]$
\end{IEEEproof}
We can now derive the gap between inner and outer bounds.
\begin{thm}{\bf Approximate capacity of the  DPC-SF-RCSI in the ``strong fading'' regime.}
\label{th:Approximate capacity of the  DPC-SF-RCSI in the strong fading regime}
The capacity $C$ of the $M$ user DPC-SF-RCSI can be attained to within a gap of $G$ defined as
\ea{
G=R^{\rm OUT}-R^{\rm IN} \leq 3+\f {\log M} M
}
\end{thm}

\begin{IEEEproof}
The gap is obtained by considering the difference between the inner bound expressions \eqref{eq:Inner bound M} and the outer bound expression in \eqref{eq:outer bound M}.
\end{IEEEproof}

The result in Th. \ref{th:Approximate capacity of the  DPC-SF-RCSI in the strong fading regime} establishes a regime in which the best coding option is
pre-code  against the interference realization at each receiver for a portion $1/M$ of the time.
This is indeed a quite pessimistic result, since this achievable scheme attains roughly $\f 1 2M \log(1+P)$ which is a fraction $M$ of the capacity for the case
without fading.
In other words, the encoder does not exploit the correlation between the received signals $Y_j$.
An intuitive explanation of this result is as follows: in the ``strong fading'' regime, the portion of the interfere which is received at the noise level at
receiver $j$ is received at the level of the intended signal at receiver $j+1$.
As an example, consider the case in which $a_j=(P+1)^j$ for which condition \eqref{eq: strong fading} hold only approximatively.
The interference $a_j S_i$ experienced at receiver $j$ can be rewritten as
\ea{
a_j S_i=S_P + S_N
}
where $S_P \sim \Ncal(0,(P+1)^j-1)$ and $S_N \sim \Ncal(0,1)$.
At receiver $j+1$ we have
\ea{
a_{j+1} S_i= \f {a_{j+1}}{a_j} (S_P + S_N) \approx (P+1) S_P + (P+1) S_N.
}
The component $S_N$ is received at the noise level at receiver $j$ but is received at the power lever $P$ at receiver $j+1$.
In other words, he interference portion which collides with the signal $X_i$ at receiver $j+1$ is observed at the noise level at receiver $j$.
This implies a ``renewal'' of how the interference collides with the signal $X_i$ at each receiver.
This makes it impossible for the receiver to pre-code across multiple realization of the fading.

Although pessimistic, this assumptions can be used to bound the capacity of more general fading realizations as well.
To do so we must pick the largest number of fading realizations with satisfy the strong fading condition in \eqref{eq: strong fading}.
\begin{lem}{\bf Outer bound for the $M$ fading values}
The capacity $C$ of the $M$ user DPC-SF-RCSI with $a_1=0$ is lower bounded as
\ea{
C \leq R^{\rm OUT} \leq \f 1 {2 K} \log (1+P)
}
where $K$ is the largest number of realization $a_j$ in $\Acal$ which satisfy the condition in \eqref{eq: strong fading}.
\end{lem}
\begin{IEEEproof}
Consider $\Acal$ and choose the largest subset of elements $\Bcal \subset \Acal$ for which \eqref{eq: strong fading}is satisfied.
Using Fano's inequality then write
\ea{
N(R -\ep)& \leq  \min_{a_j \in \Acal} I(Y_j^N; W) \leq \min_{a_j \in \Bcal} I(Y_j^N; W),
}
and apply the outer bound in Th. \ref{th:Outer bound M} for the channel output in $\Bcal$.
\end{IEEEproof}
%

\section{Conclusion}
\label{sec:Conclusion}

In this paper the capacity of the dirty paper channel with fading and  partial transmitter side information is studied.
We consider two possible scenarios: the one in which fading process takes two values and the case in which the fading process takes $M$ possible values.
For the case in which takes two values, we characterize capacity to within a constant additive gap which is small when the fading realizations are sufficiently spaced apart.
%
%
For the case with $M$ possible fading values, we show that there exists a regime in which capacity is a factor $M$ smaller with respect to the case with no fading.
%

\bibliographystyle{IEEEtran}
\bibliography{steBib1}

\begin{thebibliography}{1}
\providecommand{\url}[1]{#1}
\csname url@samestyle\endcsname
\providecommand{\newblock}{\relax}
\providecommand{\bibinfo}[2]{#2}
\providecommand{\BIBentrySTDinterwordspacing}{\spaceskip=0pt\relax}
\providecommand{\BIBentryALTinterwordstretchfactor}{4}
\providecommand{\BIBentryALTinterwordspacing}{\spaceskip=\fontdimen2\font plus
\BIBentryALTinterwordstretchfactor\fontdimen3\font minus
  \fontdimen4\font\relax}
\providecommand{\BIBforeignlanguage}[2]{{%
\expandafter\ifx\csname l@#1\endcsname\relax
\typeout{** WARNING: IEEEtran.bst: No hyphenation pattern has been}%
\typeout{** loaded for the language `#1'. Using the pattern for}%
\typeout{** the default language instead.}%
\else
\language=\csname l@#1\endcsname
\fi
#2}}
\providecommand{\BIBdecl}{\relax}
\BIBdecl

\bibitem{GelFandPinskerClassic}
S.~Gel'fand and M.~Pinsker, ``Coding for channel with random parameters,''
  \emph{Problems of control and information theory}, vol.~9, no.~1, pp. 19--31,
  1980.

\bibitem{costa1983writing}
M.~Costa, ``{Writing on dirty paper},'' \emph{IEEE Trans. Inf. Theory},
  vol.~29, no.~3, pp. 439--441, 1983.

\bibitem{zhang2007writing}
W.~Zhang, S.~Kotagiri, and J.~N. Laneman, ``Writing on dirty paper with
  resizing and its application to quasi-static fading broadcast channels,'' in
  \emph{Proc. IEEE International Symposium on Information Theory (ISIT),Nice,
  France.}, 2007, pp. 381--385.

\bibitem{cover2002duality}
T.~M. Cover and M.~Chiang, ``Duality between channel capacity and rate
  distortion with two-sided state information,'' \emph{Information Theory, IEEE
  Transactions on}, vol.~48, no.~6, pp. 1629--1638, 2002.

\bibitem{grover2007writing}
P.~Grover and A.~Sahai, ``Writing on rayleigh faded dirt: a computable upper
  bound to the outage capacity,'' in \emph{Proc. IEEE International Symposium
  on Information Theory (ISIT), Nice, France.}, 2007, pp. 2166--2170.

\bibitem{avner2010dirty}
Y.~Avner, B.~M. Zaidel, S.~Shamai, and U.~Erez, ``On the dirty paper channel
  with fading dirt,'' in \emph{Electrical and Electronics Engineers in Israel
  (IEEEI), 2010 IEEE 26th Convention of}, 2010, pp. 000\,525--000\,529.

\bibitem{piantanida2009capacity}
P.~Piantanida and S.~Shamai, ``Capacity of compound state-dependent channels
  with states known at the transmitter.''

\bibitem{LapidothCarbonCopying}
A.~Khisti, U.~Erez, A.~Lapidoth, and G.~Wornell, ``Carbon copying onto dirty
  paper,'' \emph{IEEE Trans. Inf. Theory}, vol.~53, no.~5, pp. 1814--1827, May
  2007.

\bibitem{shannon1958channels}
C.~E. Shannon, ``Channels with side information at the transmitter,'' \emph{IBM
  journal of Research and Development}, vol.~2, no.~4, pp. 289--293, 1958.

\end{thebibliography}

\onecolumn
\appendix
\subsection{Proof of Rem. \ref{rem:Generalized Channel Model} }
\label{app:Generalized Channel Model}
To show the equivalency, we show that each channel of the form of \eqref{eq:generalized channel} is in a many-to-one mapping with a channel model in \eqref{eq:DPC-SF-RCSI}.
To do so, we construct the following mapping:
\eas{
Y & = \f {Y'} {\sqrt{N'}} \\
X & = \f {X'} {\sqrt{N'}} \\
A & = A' \f{\sqrt{Q'}} {\sqrt{N'}}
}
so that $P=P'/N'$.
Since $Y$ is obtained through a linear transformations from $Y'$, we conclude that the capacity of the two models is the same.

\subsection{Proof of Lem. \ref{lem:Constant additive gap for M, small fading values}}
\label{app:Constant additive gap for M, small fading values}

The outer bound in \eqref{eq:trivial outer bound M} is obtained by providing both receivers with the state sequence.

For the inner bound, consider the Costa coding strategy in which encoder can pre-code against the average state plus fading realization
\ea{
\St = \f {\sum_{j=1}^M a_j} M S.
}
This choice attains the rates
\ea{
R_j \leq \f 12 \log\lb  \f {P+1} {1+  \f {\ep^2 P}{P+a_j^2+1}} \rb, \quad j\in [1 \ldots M],
\label{eq:constan M small}
}
at receiver $j$.
Since all the encoders must be able to decode the transmitted codeword, we must choose the smallest $R_j$ among the ones in \eqref{eq:constan M small}.
The rate $R_j$ is decreasing in $j$, so the largest gap from the trivial outer bound in \eqref{eq:trivial outer bound M} is attained at $j=1$.

\subsection{Proof of Rem. \ref{rem:capacity decreasing scaling fading} }
\label{app:capacity decreasing scaling fading}
 
 Consider the state $S$ in \eqref{eq:DPC-SF-RCSI} and assume that it is obtained as
the summation of two independent components:
\ea{
S^N=S_1^N+S_2^N,
}
for $S_{j,i} \sim \Ncal(0,Q_j), \ j \in\{1,2\}$ and $Q_1+Q_2=1$ and $i \in [1 \ldots N]$.
Consider now providing $S_2^N$ to both transmitter and receiver: this can only increase the capacity,
 since the receiver can disregard this additional information.
On the other hand, the receiver can subtract $A S_1^N$ and obtain the channel output
\ea{
Y'^N=X+A S_1^N + Z^N,
}
which corresponds to the channel where the state has power $Q_1$ in the general model in Rem. \ref{rem:Generalized Channel Model} and where $S_2^N$ has the role of common randomness.
Common randomness cannot increase capacity and therefore we conclude that capacity of the channel with state power $Q_1$ is larger that the capacity of the
channel with state power one, for any $Q_1<1$.
To re-normalize consider again the transformation in Rem. \ref{rem:Generalized Channel Model} which produces the equivalent model
\ea{
Y^N= X^N + \lb A \sqrt{Q_1}   \rb \f{S_1^N} {\sqrt{Q_1}}  + Z^N.
}

\subsection{Proof of Th. \ref{th:Outer bound 2}.}
\label{app:Outer bound 2}

The first step of the outer bound is similar to the bounding in \cite[Th. 3]{LapidothCarbonCopying} while the second passage makes use of the observation in
Rem. \ref{rem:capacity decreasing scaling fading}.

\medskip

As in \cite[Th. 3]{LapidothCarbonCopying}, we have that the capacity of the compound channel can be bounded as
\eas{
N (R -\ep)
& \leq \min_{j \in \{1,2\}}I(Y_j^N;W) \\
& \leq \f 1 2 \lb I(Y_1^N;W)  + I(Y_2^N;W)\rb \\
& = \f 1 2 \lb H(Y_1^N)+H(Y_2^N) - H(Y_1^N|W)  - H(Y_2^N| W)\rb
}
For the terms $H(Y_1^N)+H(Y_2^N)$ we have
\eas{
& H(Y_1^N)+H(Y_2^N)  \\
& \leq \f N 2 \log(P+a_1^2 + 2 a_1 \sqrt{P} +1) + \f N 2 \log(P+ a_2^2+ 2 a_2 \sqrt{P} +1)+ \log 2 \pi e \\
& \leq N \lb \f  1 2 \log(P+a_1^2 +1) + \f 1 2 \log(P+ a_2^2 +1)+ 1+ \log 2 \pi e \rb .
}
For the negative term $-H(Y_1^N|W)  - H(Y_2^N| W)$ we write
\ean{
& -H(Y_1^N|W)  - H(Y_2^N| W) \\
& \leq -H(Y_1^N,Y_2^N| W) \\
& = -H((a_2-a_1) S^N +Z_1^N-Z_2^N, Y_2^N | W) \\
}
where we have used the transformation
\ea{
\lsb
\p{Y_2-Y_1 \\
Y_2
}
\rsb
=
\lsb \p{
-1 & 1  \\
0  & 1
}
\rsb
\cdot
\lsb
\p{
Y_1 \\
Y_2
}
\rsb
}
which has determinant one.
We now continue the series of inequalities as
\ean{
& = -H((a_2-a_1) S^N +Z_1^N-Z_2^N| W) - H(Y_2^N| (a_2-a_1) S^N +Z_1^N-Z_2^N , W) \\
& \leq -H((a_2-a_1) S^N +Z_1^N-Z_2^N) - H(Y_2^N| (a_2-a_1) S^N +Z_1^N-Z_2^N , W, S^N) \\
& \leq -H((a_2-a_1) S^N +Z_1^N-Z_2^N) - H(Z_2^N| Z_1^N-Z_2^N).
}
Let now $Z_1^N \perp Z_2^N$  to obtain
\eas{
-H(Y_1^N|W)-H(Y_2^N| W) & \leq  N \lb \f 12 \log( (a_2-a_1)^2 +2) - \f 12 \log 2 \pi e \f 12 \rb.
}
The two above inequalities establish \eqref{eq:Outer bound 2 fading}.

\medskip

Since the capacity of the  channel is decreasing in $\gamma$ for $A'=\gamma A$, we can optimize the bound in  \eqref{eq:Outer bound 2 fading}:
let's rewrite as:
\ea{
\eqref{eq:Outer bound 2 fading} \leq \max_{x \in [0,1]}\f 1 2 \log\lb 1+P+ x  a_1^2 \rb +\f 12 \log \lb 1+P+x a_2^2 \rb
-\f 1 2 \log( x (a_2-a_1)^2)+1.
\label{eq:optimize Q}
}
The first derivative of  \eqref{eq:optimize Q} in
\ea{
\f{\partial \  \eqref{eq:optimize Q} }{\partial x  } = - \f 1 2 \f{(1+P)^2-x^2 a_1^2 a_2^2}{x (1+P+x a_1^2) (1+P+x a_2^2)}
}

The derivative of the expression in \eqref{eq:optimize Q} has a zero in
\ea{
x^*= \f {P+1} { a_2 a_1 }
}
this value is less than one when
\ea{
P+1 \leq a_2 a_1.
\label{eq:outer minimum}
}
The second derivative in $x^*$ is
\ea{
\lnone \f{\partial \ \eqref{eq:optimize Q} }{\partial^2 x  } \rabs_{x=x^*} = \f { a_2^3 a_1^3} {(1+P)^2 (a_1+a_2)^2}
}
which is positive defined. We therefore conclude that this is indeed a minimum when \eqref{eq:outer minimum} holds.

For the case in which  $P+1 > a_2 a_1$, we can bound \eqref{eq:Outer bound 2 fading}  as
in \eqref{eq:Outer bound 2 fading} since
\ea{
P+1 > a_2 a_1 \implies  P+1 > a_1^2
}

The result of the optimization in $\gamma$ correspond to bound in \eqref{eq:Outer bound 2 fading}.
%
%
%
%
\subsection{Proof of Th. \ref{th:inner bound for two fading values}}
\label{app:inner bound for two fading values}

Consider the transmission scheme in which the channel input, $X$, is comprised of two codewords:
\begin{itemize}
  \item a first codeword e $X_{N}^N$ ($N$ as in ``\emph{state as Noise}'') which treats the state as noise while
  \item a second codeword $X_{P}$ ($P$ as in ``\emph{Pre-coded against the state}'')  is pre-coded against the sequence $S'=[S^{N/2} \  a_2 S_{N/2+1}^N]$.
  This pre-coding offers full state pre-cancellation half of the time while at each of the decoders.
  For the remaining time, only partial state pre-coding is possible.
\end{itemize}

We consider, in particular, the assignment
\eas{
X_{N} & \sim \Ncal(0, \al P) \\
X_{P} & \sim \Ncal(0, \alb P) \\
U_{P} &= X_{P}+ \f{\alb P}{\alb P+1}S'.
}{\label{eq:assigment inner bound 2}}
where $\al \in [0,1], \ \alb=1-\al$.

The codeword that treats the state as noise can be decoded at both receivers when
\eas{
R_N  &\leq \min \lcb I(Y_1;X_N), I(Y_2;X_N) \rcb  \\
     & =   \min \lcb \f 12 \log\lb 1 + \f{ \alb P} {a_1^2+\al P +1}  \rb, \f 12 \log\lb 1 + \f{ \alb P } {a_2^2+\al P +1} \rb \rcb   \\
     & =   \f 12 \log \lb 1 + \f{ \alb P} {a_2^2+\al P +1} \rb
}{}

The codeword $X_{N}$ is decoded first  and removed from the channel output and, successively, the codeword $X_{P}^N$:
for the first $N/2$ transmission, the codeword can be decoded by the first receiver, while for the second $N/2$ transmissions it can be decoded by the second receiver.
%

This assignment attains
\ea{
R = \f 14 \log (1+ \al P) +  \f 12 \log\lb 1 + \f{\alb P }{a_2^2+\al P+1}\rb
\label{eq: achievable 2 values}
}
%
%
%

We now optimize the achievable scheme through an appropriate choice of $\al$: the expression

\ea{
\f {\partial \ \eqref{eq: achievable 2 values}} {\partial \al } = - \f 1 4  \f {P (\al P+1 -a_2^2) }{(a_2^2+\al P+1)(1+\al P)}
}
which has a zero in
\ea{
\al^*= \f{a_2^2-1}{P}
}
which is less than one when
\ea{
a_2^2 \leq P+1,
\label{eq:inner bound minimum}
}
while it is positive for
\ea{
a_2^2 >1
}
The second derivative in $x^*$  is
\ea{
\lnone \f {\partial \ \eqref{eq: achievable 2 values}} {\partial^2 \al } \rabs_{x=x^*} = - \f 1 8 \f{P^2} {a_2^4 },
}
so we conclude that this is indeed a maximum.

We then conclude that optimal choice of $\al$ is
\ea{
\al^* =\lcb
\p{
0 & a_2^2 \leq 1 \\
\f{a_2^2-1}{P} &  a_2^2 \leq P+1 \\
1 & a_2^2 > P+1 \\
}
\rnone
}
which produces the achievable rates in \eqref{eq:inner bound for two fading values}.
%
%

\subsection{Proof of Th. \ref{th:Approximate Capacity 2}}
\label{app:Approximate Capacity 2}

Let's begin by considering the case of in which $P$, $a_1$ or $a_2$ are small.

\medskip
\emph{Small $P$}
\smallskip

A trivial outer bound to the capacity of the DPC-SF-RCSI is
\ea{
R \leq \f 12 \log(1+P).
\label{eq:trivial outer}
}
If  $P<1$, then this outer bound is smaller than $1/2$ which means that capacity can be achieved to within $1/2$ a bit without any transmission taking place.

\medskip
\emph{Small $a_2$}
\smallskip

A trivial inner bound is to treat the state as noise, which achieves
\ea{
R \leq \f 12 \log \lb 1 + \f P {a_2^2+1}\rb.
}
If $a_2^2 <1$, this strategy achieves
\ea{
R \leq \f  12 \log \lb 1 + \f P2 \rb
}
which is to within half  a bit from the trivial outer bound of \eqref{eq:trivial outer}.

\medskip

Another favorable case is the one in which $a_2-a_1$ is small: in this case \ref{lem:Constant additive gap for two, small fading values} applies.

\medskip
\emph{Small $a_2-a_1$}
\smallskip

If $a_2-a_1 \leq 4$, then the gap between inner and outer bound is less than $1$ bit/s

\bigskip

We  now focus on determining a constant gap is the following cases:
\begin{itemize}
  \item Case I: $a_1 a_2 \geq P+1$, $a_2> \sqrt{P+1}$,
  \item Case II: $1< a_2^2 < P+1$,
  \item Case III: $a_1 a_2 < P+1$, $a_2> \sqrt{P+1}$.
\end{itemize}

\medskip
\noindent
\emph{Case I: $a_1 a_2 \geq P+1$, $a_2> \sqrt{P+1}$}
\smallskip

In this case the outer bound is
\ea{
R^{\rm OUT}=\f 14 \log (P+1)+ \f 14 \log \f {(a_2+a_1)^2}{(a_2-a_1)^2} +1
}
while the inner bound is
\ea{
R^{\rm IN}=\f 14 \log (P+1)
}
The gap between inner and outer bound is therefore
\ea{
G_I=\f 14 \log \f {(a_2+a_1)^2}{(a_2-a_1)^2}+1
}

\medskip
\noindent
\emph{Case II: $ 1< a_2^2 < P+1$}
\smallskip

In this case the outer bound is
\ea{
R^{\rm OUT}=\f 14 \log (1+P) + \f 14 \log\lb \f{P+1}{a_2^2}\rb +\f 14 \log \f {a_2^2}{(a_2-a_1)^2}+ \f 3 2
}
while the inner bound is
\ea{
R^{\rm IN}=\f 12 \log \lb 1+ P+ a_2^2 \rb - \f 14 \log (a_2^2)-1/2
}
The gap between inner and outer bound is therefore
\ea{
G_{II}=\f 14 \log \f {a_2^2}{(a_2-a_1)^2}+2
}

\medskip
\noindent
\emph{Case III: $a_1 a_2 < P+1$, $a_2> \sqrt{P+1}$}
\smallskip

In this case the outer bound is
\ea{
R^{\rm OUT}=\f 14 \log (1+P) + \f 14 \log\lb P+1 +a_2^2 \rb -\f 14 \log (a_2-a_1)^2+ \f 3 2
}
which can be further rewritten as
\ea{
R^{\rm OUT}=\f 14 \log (1+P) + \f 14 \log\lb \f {a_2^2}{(a_2-a_1)^2} \rb     + 2
}
which can be achieved with binning.

Combining the three gaps we obtain the desired result.

\subsection{Proof of Th. \ref{th:Approximate capacity of the  DPC-SF-RCSI in the strong fading regime}}
\label{app:Approximate capacity of the  DPC-SF-RCSI in the strong fading regime}

Consider now the following outer bound
\ean{
N(R -\ep)& \leq  \min_{j \in \{1 \ldots M\}} \lcb I(Y_j^N; W) \rcb  \\
& \leq \f 1 M \lb \sum_{j=1}^M I(Y_j^N; W) \rb
}
For each term $I(Y_j^N; W)$ for $2 \leq $we provide the side information $V_j^N$ for $j\geq 3$ defined as
\ea{
V_j^N & = \lsb V_{j-1}^N \ S_{j-1}^N + Z_{j-1}^N-Z_1^N \rsb
}
with this side information, and for $V_1^N=V_2^N=\emptyset$, we write
\eas{
\sum_{j=1}^M I(Y_j^N; W) & \leq \sum_{j=1}^M I(Y_j^N, V_j^N ; W) \\
& = \sum_{j=1}^M I(Y_j^N ; W| V_j^N)  \\
& = \sum_{j=1}^M \lb  H(Y_j^N | V_j^N)-H(Y_j^N | W, V_j^N) \rb
\label{eq:pos minus neg }
}
For the term
\ean{
H(Y_1^N | V_1^N)& =H(X+Z_1)  \\
& \leq \f 12 \log(2 \pi(P+1))
}
and similarly
\ean{
H(Y_2^N | V_2^N)=H(X+a_2 S + Z_2) & \leq \f 12 \log 2 \pi(P+a_2^2 + 2 a_2 \sqrt{P}+1) \\
& \leq \f 12 \log 2 \pi(P+a_2^2 +1)
}

For the remaining terms and considering independent noise terms $Z_j$, we have:
\ean{
H(Y_q^N | V_q^N)
& = H(X+a_q S + Z_q | a_2 S +Z_2, a_3 S + Z_3, \ldots, a_{j-1} S + Z_q) \\
& \leq H(X+a_q S + Z_q | S + \Zt)\\
}
for
\ea{
\Zt = \f {\sum_{j=1}^{q-1} a_j Z_j}{\sum_{j=1}^{q-1} a_j^2} \sim \Ncal \lb 0, \f {1} {\sum_{j=1}^{q-1} a_j^2} \rb
}
The last passage follows from the maximal ration combining principle.
We continue the sequence of inequalities as
\ean{
H(Y_q^N | V_q^N) & \leq H(X+a_q S + Z_q | S + \Zt)\\
& = H(X+ Z_q - a_q  \Zt | S + \Zt )\\
& \leq H(X+ Z_q - a_q  \Zt ) \\
& \leq \f 12 \log 2 \pi e \lb P+ 1 + \f {a_q^2} {\sum_{j=1}^{q-1} a_j^2} \rb
}

We now establish a recursion for the negative entropy terms: the recursion step initiates as follows:
\ean{
-H(Y_1|W) - H(Y_2|W) \leq  H(Y_1,Y_2| W)
}
Consider now the transformation
\ea{
\lsb \p{
X-Y \\
Y
} \rsb =
\lsb \p{
1 & -1 \\
0 & 1
} \rsb
\cdot
\lsb \p{
X \\
Y
} \rsb
}
whose Jacobian has determinant one.
With this transformation we write
\ean{
-H(Y_1^N,Y_2^N-Y_1^N| W) & = -H(a_2 S^N +Z_2^N) - H(Y_1^N | W , a_2 S^N +Z_2^N-Z_1^N) \\
&= - \f N 2 \log 2 \pi e \lb a_2^2 + 1\rb  - H(Y_1^N | W , a_2 S^N +Z_2^N-Z_1^N) \\
& = - \f N 2 \log 2 \pi e \lb a_2^2 + 1\rb  - H(X^N -Z_1^N| W , a_2 S^N +Z_2^N-Z_1^N)
}
Let now
\ea{
T_q = -\sum_{j=1}^{q-1} H(Y_j^N| W, V_j^N)
}
so that
\ea{
T_2 \leq \f N 2 \log 2 \pi e \lb a_2^2 + 2\rb  - H(Y_1^N | W , a_2 S^N +Z_2^N-Z_1^N)
}

For the term $T_3$ we write
\ean{
T_3 & =  T_2 - H(Y_3 | W , a_2 N + Z_2^N - Z_1^N)\\
& \leq \f N 2 \log 2 \pi e \lb a_2^2 + 1\rb  - H(X_1^N -Z_1^N| W , a_2 S^N +Z_2^N) - H(Y_3 | W , a_2 S^N + Z_2^N-Z_1^N) \\
& = \f N 2 \log 2 \pi e \lb a_2^2 + 1\rb - H(X^N-Z_1^N, Y_3 | W , a_2 S^N +Z_2^N-Z_1^N)
}
With the same transformation as above we have
\ean{
& H(X^N-Z_1^N, Y_3^N | W , a_2 S^N +Z_2^N)  \\
& = H(X^N-Z_1^N, Y_3^N-Y_1^N | W , a_2 S^N +Z_2^N-Z_1^N ) \\
& = H(a_3 S^N + Z_3^N - Z_1^N| a_2 S^N +Z_2^N-Z_1^N) + H(X^N-Z_1^N | W , a_2 S^N +Z_2^N-Z_1^N , a_3 S^N + Z_3^N - Z_1^N ).
}

Combining the above two equations we have
\ean{
T_3 \leq - \sum_{q=1}^3  H(V_q^N|V_{q-1}^N) -  H(Y_1^N | W ,V_j^N).
}

Continuing this recursion we see that the $j^{\rm th}$ term $T_j$ can be bounded as
\ean{
T_q & = T_{q-1} - H(Y_j | W , a_2 S^N + Z_2^N - Z_1^N, \ldots a_{j-1} S^N+Z_{j-1}^N) \\
& \leq  - \sum_{j=1}^{q-1} H(V_j^N | V_{j-1}^N)-H(X^N-Z_1^N|V_{q})
}
For the final step of the recursion we write
\ean{
T_M & \leq - \sum_{j=1}^M H(V_j^N | V_{j-1}^N)-H(Y_1^N|V_j) - H(X^N-Z_1^N| V_M^N)  \\
     & \leq \sum_{j=1}^M H(V_j^N | V_{j-1}^N)-H(Y_1^N|V_j) - H(X^N-Z_1^N| V_M^N, X^M)  \\
     & = - \sum_{q}^M H(V_q^N | V_{q-1}^N)-H(Y_1^N|V_j) - H(Z_1^N|a_2 N + Z_2 -Z_1, \ldots a_M N + Z_M -Z_1)
}

We are now left to evaluate the terms $H(V_q^N | V_{q-1}^N)$  and the term $H(Z_1^N|a_2 N + Z_2 -Z_1, \ldots a_M N + Z_M -Z_1)$.
For the term $H(V_q^N | V_{q-1}^N)$ we proceed as follows:
\ean{
& -H(V_q^N | V_{q-1}^N)\\
& \leq - H(V_q^N | V_{q-1}^N, Z_1) \\
& = H( a_q S^N + Z_q^N | a_2 S^N+Z_2^N, \ldots a_{q-1} S^N + Z_q^N)\\
& = H \lb a_q S^N + Z_q^N | \lb \f {\sum_{j=1}^q a_j (a_j S_j + Z_j)}{\sum_{j=1}^q a_j^2 }\rb   \rb \\
& = H \lb a_q S^N + Z_q^N | S^N+\Zt^N \rb
}
where the last passage follows from the maximal ratio combining principle and by letting
\ea{
\Zt = \f {\sum_{j=1}^q a_j Z_j}{\sum_{j=1}^q a_j^2} \sim \Ncal \lb 0, \f {1} {\sum_{j=1}^N a_j^2} \rb
}
We therefore have
\ea{
-H(V_q^N | V_{q-1}^N) \leq \f 12 \log 2 \pi e \lb 1+ \f{a_q^2}{1+ \sum_{j=1}^N a_j^2}\rb
}

In a similar fashion, we write
\ean{
& -H(Z_1^N|a_2 S^N + Z_2 -Z_1, \ldots , a_M S^N + Z_M -Z_1) \\
& \leq -H(Z_1^N|a_2 S^N + Z_2 -Z_1, \ldots , a_M S^N + Z_M -Z_1, S^N) \\
& \leq -H(Z_1^N|Z_2 -Z_1, \ldots , Z_M -Z_1) \\
& \leq -H \lb Z_1^N | Z_1^N - \f{\sum_{j=2}^M  Z_j} {M-1} \rb \\
& = \f 12 \log 2 \pi e \log \lb 1 - \f 1 {1 + \f 1 {M-1}}\rb \\
& = \f 12 \log 2 \pi e \log \lb \f 1 M \rb
}

We now wish to return to the expression in \eqref{eq:pos minus neg }  and show that
\ea{
H(Y_j^N|V_j)-H(V_j^N | V_{j-1}^N)
}
is bounded under the conditions in \eqref{eq: strong fading} and $j\geq 2$
\ean{
& H(Y_j^N|V_j)-H(V_j^N | V_{j-1}^N) \\
& \leq \f 12 \log \lb \f{ P+1 + \f {a_j^2}{\sum_{j=1}^{q-1} {a_j^2}+1}} {1+ \f{a_j^2}{1+ \sum_{j=1}^N a_j^2} }\rb  \\
& = \f 12 \log \lb \f{ (\sum_{j=1}^{q-1} a_j^2+1)(P+1) + a_j^2} {\sum_{j=1}^{q-1} a_j^2+1 +a_j^2} \rb  \\
& = \f 12 \log \lb \f{ (\sum_{j=1}^{q-1} a_j^2+1)(P+1) + a_j^2} {\sum_{j=1}^{q} a_j^2+1} \rb  \\
}
When  condition \eqref{eq: strong fading} holds, we have therefore have
\ean{
& H(Y_j^N|V_j)-H(V_j^N | V_{j-1}^N) \\
& \leq \f 12 \log \lb 1+\f{3a_q^2} {\sum_{j=1}^{q} a_j^2+1} \rb  \\
& \leq 2.
}

The only remaining terms are  $H(Y_1)-H(Y_1|V_M)$ for which we have
\ean{
& H(Y_1)-H(Y_1|V_M)  \\
& \leq H(Y_1)-H(Y_1|V_M, S^N,X^N) \\
& = H(X^N+Z_1^N)+\log 2 \pi e M \\
& = \f 12 \log (1+P) +\log 2 \pi e M
}
This concludes the proof.

\end{document}